\documentclass[preprint,12pt]{elsarticle}

\usepackage{amssymb}

\journal{Physics Letters B}

\begin{document}

\begin{frontmatter}



\title{Light WIMP search in XMASS}



\author[ICRR,IPMU]{K.~Abe}
\author[ICRR]{K.~Hieda}
\author[ICRR,IPMU]{K.~Hiraide}
\author[ICRR]{S.~Hirano}
\author[ICRR,IPMU]{Y.~Kishimoto}
\author[ICRR,IPMU]{K.~Kobayashi}
\author[ICRR,IPMU]{S.~Moriyama}
\author[ICRR]{K.~Nakagawa}
\author[ICRR,IPMU]{M.~Nakahata}
\author[ICRR,IPMU]{H.~Ogawa}
\author[ICRR]{N.~Oka}
\author[ICRR,IPMU]{H.~Sekiya}
\author[ICRR]{A.~Shinozaki}
\author[ICRR,IPMU]{Y.~Suzuki}
\author[ICRR,IPMU]{A.~Takeda}
\author[ICRR]{O.~Takachio}
\author[ICRR]{K.~Ueshima\fnref{tohoku}}
\author[ICRR]{D.~Umemoto}
\author[ICRR,IPMU]{M.~Yamashita}
\author[ICRR]{B.~S.~Yang}

\author[GIFU]{S.~Tasaka}

\author[IPMU]{J.~Liu}
\author[IPMU]{K.~Martens}

\author[KOBE]{K.~Hosokawa}
\author[KOBE]{K.~Miuchi}
\author[KOBE]{A.~Murata}
\author[KOBE]{Y.~Onishi}
\author[KOBE]{Y.~Otsuka}
\author[KOBE,IPMU]{Y.~Takeuchi}

\author[KRISS]{Y.~H.~Kim}
\author[KRISS]{K.~B.~Lee}
\author[KRISS]{M.~K.~Lee}
\author[KRISS]{J.~S.~Lee}

\author[MIYA]{Y.~Fukuda}

\author[NAGOYA,KMS]{Y.~Itow}
\author[NAGOYA]{K.~Masuda}
\author[NAGOYA]{Y.~Nishitani}
\author[NAGOYA]{H.~Takiya}
\author[NAGOYA]{H.~Uchida}

\author[SEJONG]{N.~Y.~Kim}
\author[SEJONG]{Y.~D.~Kim}

\author[TOKAI1]{F.~Kusaba}
\author[TOKAI2]{D.~Motoki\fnref{tohoku}}
\author[TOKAI1]{K.~Nishijima}

\author[YNU]{K.~Fujii}
\author[YNU]{I.~Murayama}
\author[YNU]{S.~Nakamura}

\address[ICRR]{Kamioka Observatory, Institute for Cosmic Ray Research,
  the University of Tokyo, Higashi-Mozumi, Kamioka, Hida, Gifu, 506-1205, Japan}
\address[GIFU]{Information and multimedia center, Gifu University, Gifu 501-1193, Japan}
\address[IPMU]{Kavli Institute for the Physics and Matematics of the Universe,
  the University of Tokyo, Kashiwa, Chiba, 277-8582, Japan}
\address[KMS]{Kobayashi-Maskawa Institute for the Origin of Particles and the Universe, 
Nagoya University, Furo-cho, Chikusa-ku, Nagoya, Aichi, 464-8602, Japan.}
\address[KOBE]{Department of Physics, Kobe University, Kobe, Hyogo 657-8501, Japan}
\address[KRISS]{Korea Research Institute of Standards and Science, Daejeon 305-340, South Korea}
\address[MIYA]{Department of Physics, Miyagi University of Education, Sendai, Miyagi 980-0845, Japan}
\address[NAGOYA]{Solar Terrestrial Environment Laboratory, Nagoya University, 
Nagoya, Aichi 464-8602, Japan}
\address[SEJONG]{Department of Physics, Sejong University, Seoul 143-747, South Korea}
\address[TOKAI1]{Department of Physics, Tokai University, Hiratsuka,
  Kanagawa 259-1292, Japan}
\address[TOKAI2]{School of Science and Technology, Tokai University, Hiratsuka,
  Kanagawa 259-1292, Japan}
\address[YNU]{Department of Physics, Faculty of Engineering, Yokohama National University, Yokohama, Kanagawa 240-8501, Japan}

\fntext[tohoku]{Now at Research Center for Neutrino Science, Tohoku University, Sendai 980-8578, Japan}

\begin{abstract}

A search for light dark matter using low-threshold data from the single phase 
liquid xenon scintillation detector XMASS, has been conducted. 
Using the entire 835\,kg inner volume as target, the analysis threshold can be lowered 
to 0.3\,keVee (electron-equivalent) to search for light dark matter. 
With low-threshold data corresponding to a 5591.4\,kg$\cdot$day exposure of the detector and
without discriminating between nuclear-recoil and electronic events, XMASS 
excludes part of the parameter space favored by other experiments.

\end{abstract}

\begin{keyword}
Dark Matter\sep WIMP\sep xenon

\end{keyword}

\end{frontmatter}


\section{Introduction}
\label{intro}
There is substantial evidence that the universe is composed of dark matter~\cite{PDG}.
Among the most plausible dark matter candidates, Weakly Interacting Massive Particles (WIMPs),
which are expected to couple to ordinary matter primarily through the weak force, can 
be detected directly through observation of nuclear recoils produced 
in their elastic scattering interactions with detector nuclei~\cite{goodman}. 
Many theories of physics beyond the Standard Model predict stable WIMPs 
with supersymmetric models still among the most popular~\cite{jungman}.
In the constrained Minimal Supersymmetric Standard Model, 
the lightest supersymmetric particle is favored as a WIMP candidate and has an expected 
mass of $M_{\chi} > 100$\,$\mbox{GeV}$.
However, some experiments indicate a possible WIMP signal~\cite{dama,cogent,cresst} 
with a lighter mass of $\sim 10$\,$\mbox{GeV}$ and with a spin-independent cross section of
the order of $\sim 10^{-40} \mbox{cm}^{2}$.
These positive signals have come predominantly from experiments without the capability 
to discriminate between electromagnetic and nuclear recoils, while other experiments 
that have this ability
have excluded light WIMPs at 
these cross sections~\cite{xenon,cdms,edelweiss}. 
This Letter presents a search for light WIMPs using a 5591.4\,kg$\cdot$day exposure of the 
XMASS experiment without nuclear recoil discrimination. 


\section{XMASS detector}
\label{detector}
The XMASS experiment is located underground in the Kamioka Observatory at a depth of 
2700\,m.w.e. and has been designed to search for WIMP interactions using the 
self-shielding properties of liquid xenon in an ultra low radioactivity environment~\cite{suzuki}.  
XMASS is a single phase liquid xenon scintillator detector containing 1050\,kg of Xe in an OFHC copper vessel. 
Xenon scintillation light is collected by 630 hexagonal and 12 cylindrical 
inward-pointing Hamamatsu R10789 series photomultiplier tubes (PMTs)
arranged on an 80\,cm diameter pentakis-dodecahedron support structure within the vessel to give a total 
photocathode coverage of 62.4\,\%. 

These PMTs view an active target region containing 835\,kg of liquid xenon.
In order to  monitor the PMT stability and measure the trigger efficiency,
eight blue LEDs with Teflon diffusers are mounted to the support structure. 
There are six LEDs arranged along the equator and one each at the top and the bottom of the pentakis-dodecahedron.
To shield the scintillator volume from external gammas, neutrons, and muon-induced backgrounds, 
the copper vessel is placed at the center of a $\phi$\,10\,m $\times$ 11\,m 
cylindrical tank filled with pure water.  This volume is viewed by 
72 Hamamatsu R3600 20-inch PMTs to provide both an active muon veto and passive shielding against these backgrounds. 
This is the first water Cherenkov shield used in a dark matter search experiment.
To perform energy and position reconstruction calibrations a portal has been prepared along 
the central vertical axis ($z$-axis) of the PMT support structure through which an OFHC copper rod 
can be inserted directly into the target volume. Thin cylindrical calibration sources  
containing either of $^{55}$Fe, $^{57}$Co, $^{109}$Cd, or $^{241}$Am are placed at the tip of this 
rod to perform detector calibrations. 
A more detailed description of the XMASS detector is presented in~\cite{xmass_det}.

PMT signals are passed though preamplifiers with a voltage gain factor of 11 before being processed 
by Analog-Timing-Modules (ATMs)~\cite{sk}.  
These modules combine the functions of typical ADC and TDC modules, recording both the integrated charge and the arrival time of each PMT signal.  
For each PMT channel the discriminator threshold is set to $-5$\,mV, which corresponds to $0.2$ photoelectrons (p.e.). 
When a PMT signal is above threshold a ``hit'' is registered on the module.
A global trigger is generated based on the number of hit PMTs within a 200\,ns window.

A complete XMASS detector Monte Carlo (MC) simulation package based on Geant4~\cite{geant4}
and including readout electronics has been developed~\cite{xmass_det} and used for this analysis.
The simulation has been tuned using calibration data and the optical 
properties of the liquid xenon have been determined empirically using 
data taken at nine points along the $z$-axis: $z=\pm40$\,cm, $\pm30$\,cm, $\pm20$\,cm, $\pm10$\,cm, 0\,cm.

\section{Data and Event Selection}
\label{data}
The data used for this analysis, corresponding to 6.70 days 
of livetime, was taken in February 2012 with a low trigger threshold of four PMT 
hits. 
Using 122\,keV gammas from the $^{57}$Co calibration source the xenon light yield 
was found to be 14.7\,photoelectrons/keVee. 
This large light yield allows the analysis threshold to be lowered sufficiently for sensitivity to low mass WIMPs. 
In order to achieve optimal sensitivity, the entire detector volume is used
because fiducialization is increasingly difficult at these low energies.

\begin{figure}[p]
\begin{center}
\includegraphics[width=8.5cm]{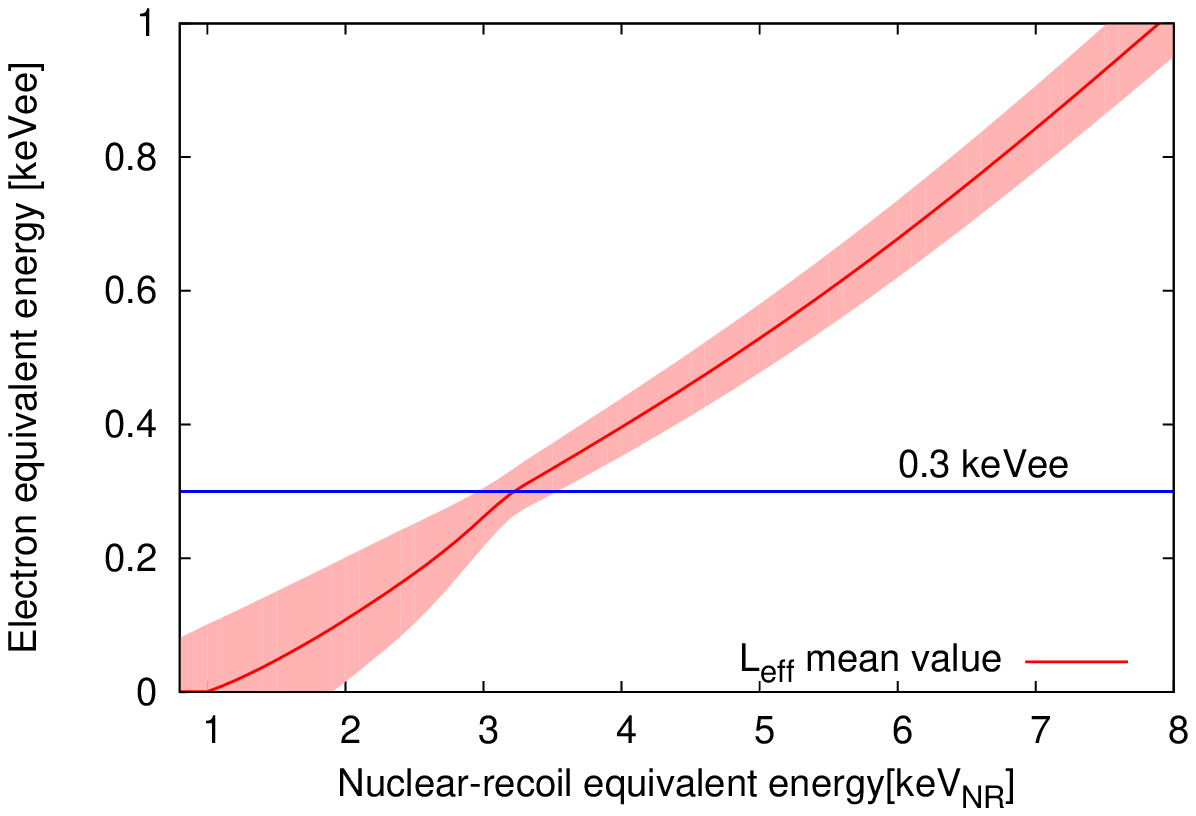}
\caption{The relationship between the electron equivalent energy and the nuclear-recoil equivalent energy 
         based on the $\mathcal{L}_{\rm eff}$ method of Ref.~\cite{Leff}.
         The one-$\sigma$ uncertainty and the 0.3\,keVee analysis threshold are also shown.}

\label{fig:Leff}
\end{center}
\end{figure}

\begin{figure}[p]
\begin{center}
\includegraphics[width=8.5cm]{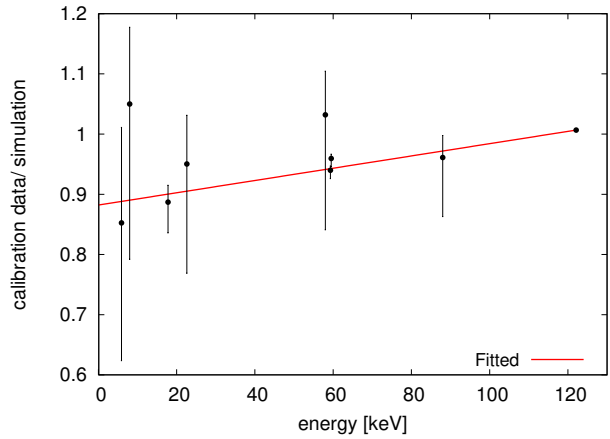}
\caption{Energy scale comparison between calibration data and simulation results. 
         Calibrations were conducted  at 122\,keV ($^{57}$Co), 88\,keV, 
         58\,keV, 22\,keV, 8\,keV ($^{109}$Cd), 59.5\,keV, 
         17.8\,keV ($^{241}$Am), and  5.9\,keV ($^{55}$Fe).
         The error bars in the figure stem from uncertainties in the surface reflection 
         properties of the calibration source housing  within the simulation.
         }
\label{fig:response}
\end{center}
\end{figure}

In the analysis presented below, the nuclear-recoil equivalent energy ($E_{\rm NR}$)
is determined by conversion from electron equivalent energy ($E_{\rm ee}$)
using the scintillation efficiency, $\mathcal{L}_{\rm eff}$, 
for nuclear recoils relative to that of 122\,keV gammas at zero electric field. 
The treatment of the energy dependence of $\mathcal{L}_{\rm eff}$ and its uncertainty is the same as in Ref.~\cite{Leff}.
The resulting functional dependence of 
 $E_{\rm NR}$ on $E_{\rm ee}$
with its one-$\sigma$ uncertainty is shown in Fig.~\ref{fig:Leff}. 
To illustrate the impact of this uncertainty on the present analysis, 
the 0.3\,keVee analysis threshold is also shown in the figure. 
In the $\mathcal{L}_{\rm eff}$ framework, though the absolute energy scale is determined at 122\,keVee,
dark matter signals are expected to appear at lower energies.
Lower energy calibration samples are used to evaluate the accuracy of the detector simulation 
at these lower energies. 
Fig.~\ref{fig:response} shows the ratio of the total number of observed photoelectrons relative to the 
simulation's prediction as a function of the electron equivalent energy for several calibration samples. 
The apparent deviation of the simulation from the data 
reflects not only imperfection in the modeling of the detector environment  
but also the non-linear response of electron-mediated events in liquid xenon~\cite{ele_non_lin}.
Since the 5.9\,keV X-ray from $^{55}$Fe is the lowest energy calibration point, 
the response at lower energies is extrapolated using a linear fit through all calibration energies. 
In the analysis the fitting error from this procedure, 1.3\,\% at 0.3\,keVee, is treated 
as a systematic effect on the energy scale. 
However, the effect of this uncertainty is small relative to that induced by the 
uncertainty in $\mathcal{L}_{\rm eff}$, which is 13~\% at the same energy.

\begin{figure}[p]
\begin{center}
\includegraphics[width=8.5cm]{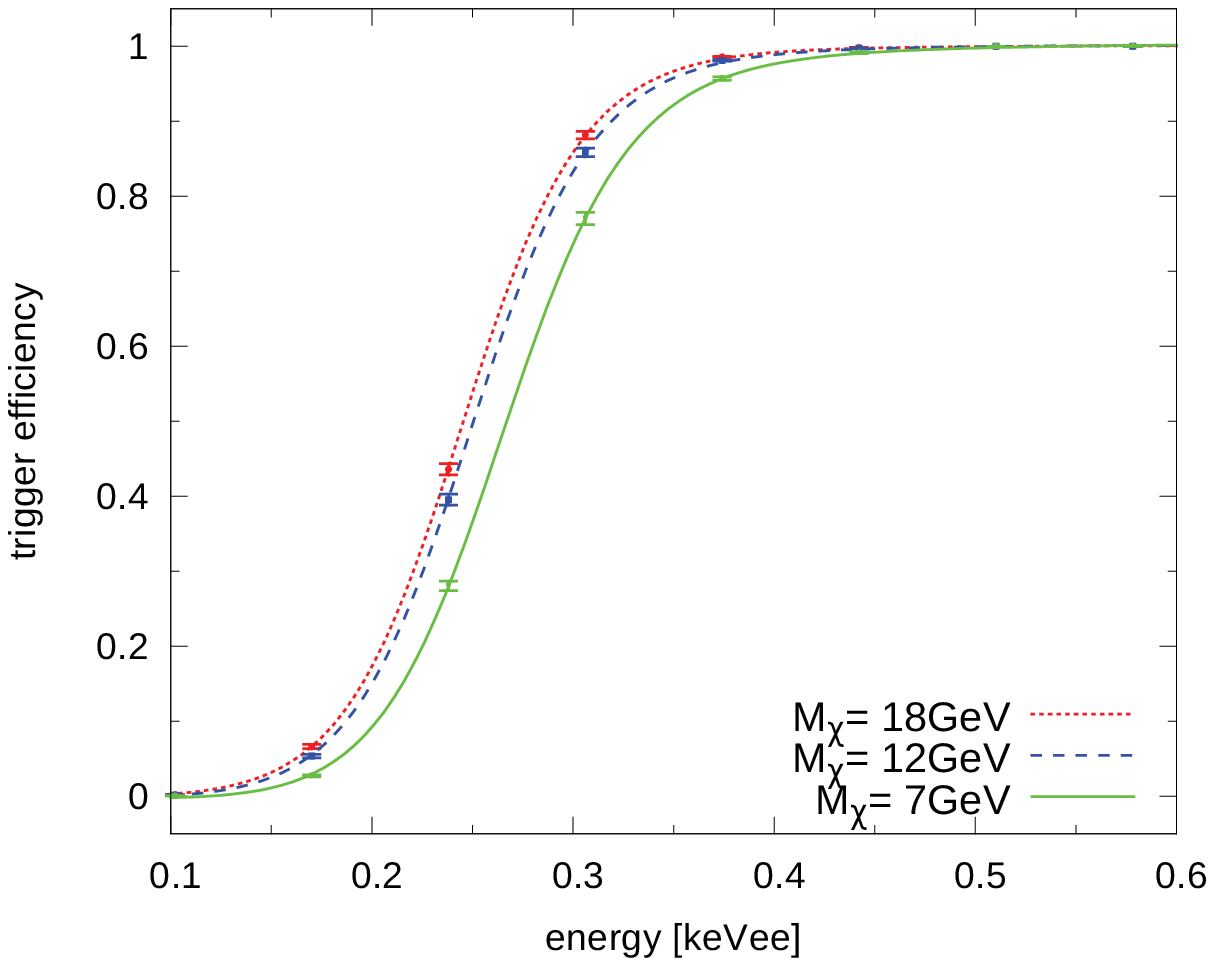}
\caption{ Expected trigger efficiencies for 7\,GeV, 12\,GeV, and 18\,GeV WIMPs 
          derived from LED data in the XMASS detector. 
        }
\label{fig:Trig}
\end{center}
\end{figure}

The trigger efficiency as a function of the 
total number of photoelectrons near the trigger threshold is smeared by the response of the electronics. 
Moreover, the trigger efficiency depends upon the WIMP mass since the distribution of the expected number of 
hits is a function of that mass. 
For this reason the trigger efficiency near threshold is derived using LED data taken at
single photoelectron level intensities in conjunction with the detector simulation. 
During the measurement LEDs are flashed at 100\,Hz and each flash of the LED is recorded regardless of the number of PMT hits observed.  
Based on this information the distribution of the number of hit PMTs, the nhit distribution, is formed for each LED.
Using the detector simulation the same distribution is made for each WIMP mass 
(The astrophysical parameters used in this simulation are described in Section~\ref{results}).
Each LED event is then reweighted based on the ratio of the nhit distributions so that 
the number of expected photoelectrons, and hence energy, for a WIMP signal can be derived 
from the LED data. 
Comparison of the number of LED flashes with the number of global triggers issued by the ATMs is then 
used to compute the expected WIMP trigger efficiency as a function of the observed energy.  
The resulting efficiency curves for representative masses from the allowed region of~\cite{cogent} are shown in
Fig.~\ref{fig:Trig}. 
The analysis threshold is chosen as the energy at which the trigger efficiency 
is greater than 50\,\% for 5\,$\mbox{GeV}$ WIMPs and corresponds to 0.3\,keVee.
The error bars in the figure illustrate that the trigger efficiency varies somewhat based upon which LED is used.
This variation reflects intrinsic differences in the LEDs themselves as well as a 
possible position dependence of the detector response and is therefore treated as 
a systematic error in the analysis. A 1.5\% error is determined at 0.3\,keVee.
Using only the predicted nhit distribution from the WIMP simulation, the efficiency curve's shape 
is well reproduced within errors. 

\begin{figure}[p]
\begin{center}
\includegraphics[width=6.5cm]{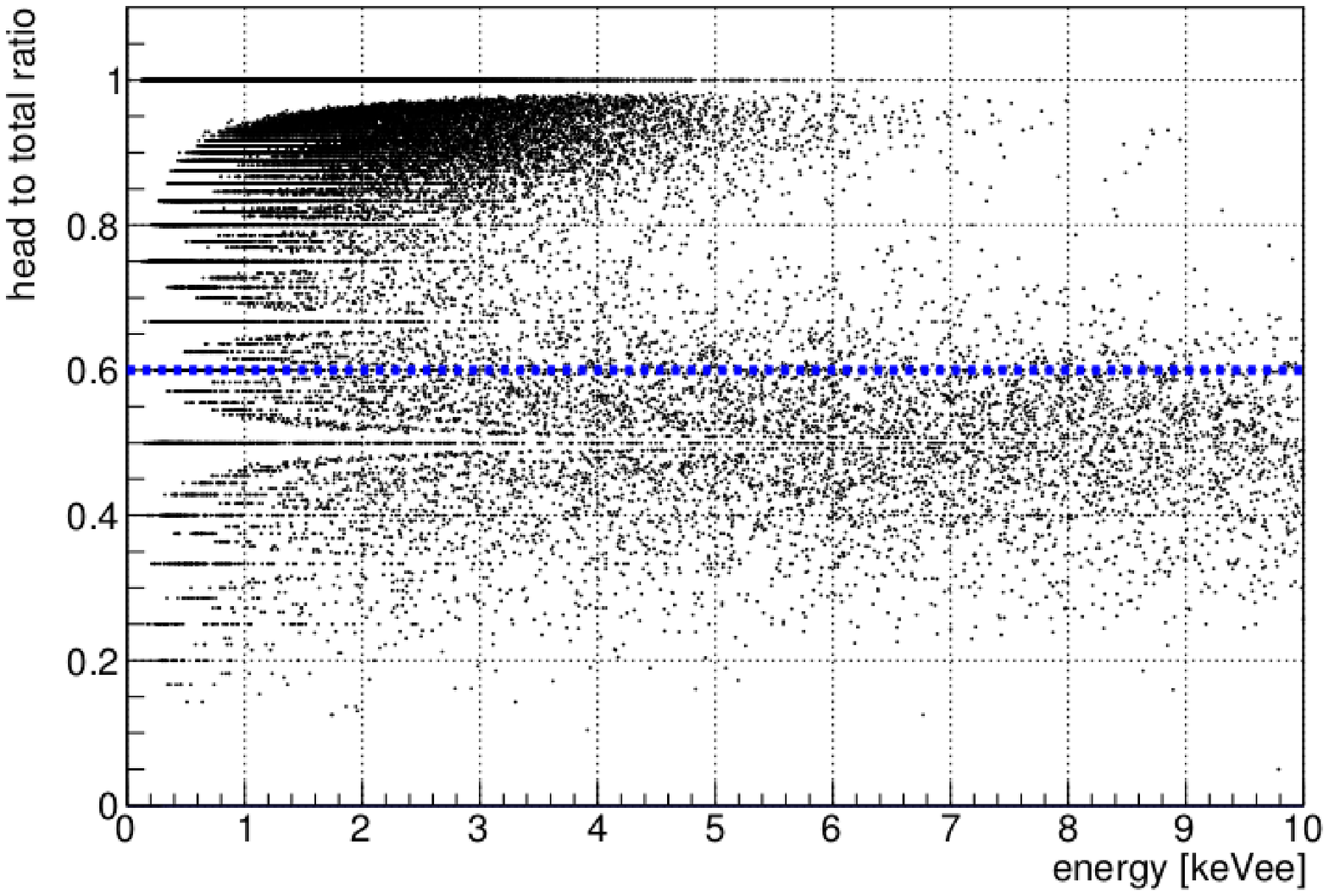}
\includegraphics[width=6.5cm]{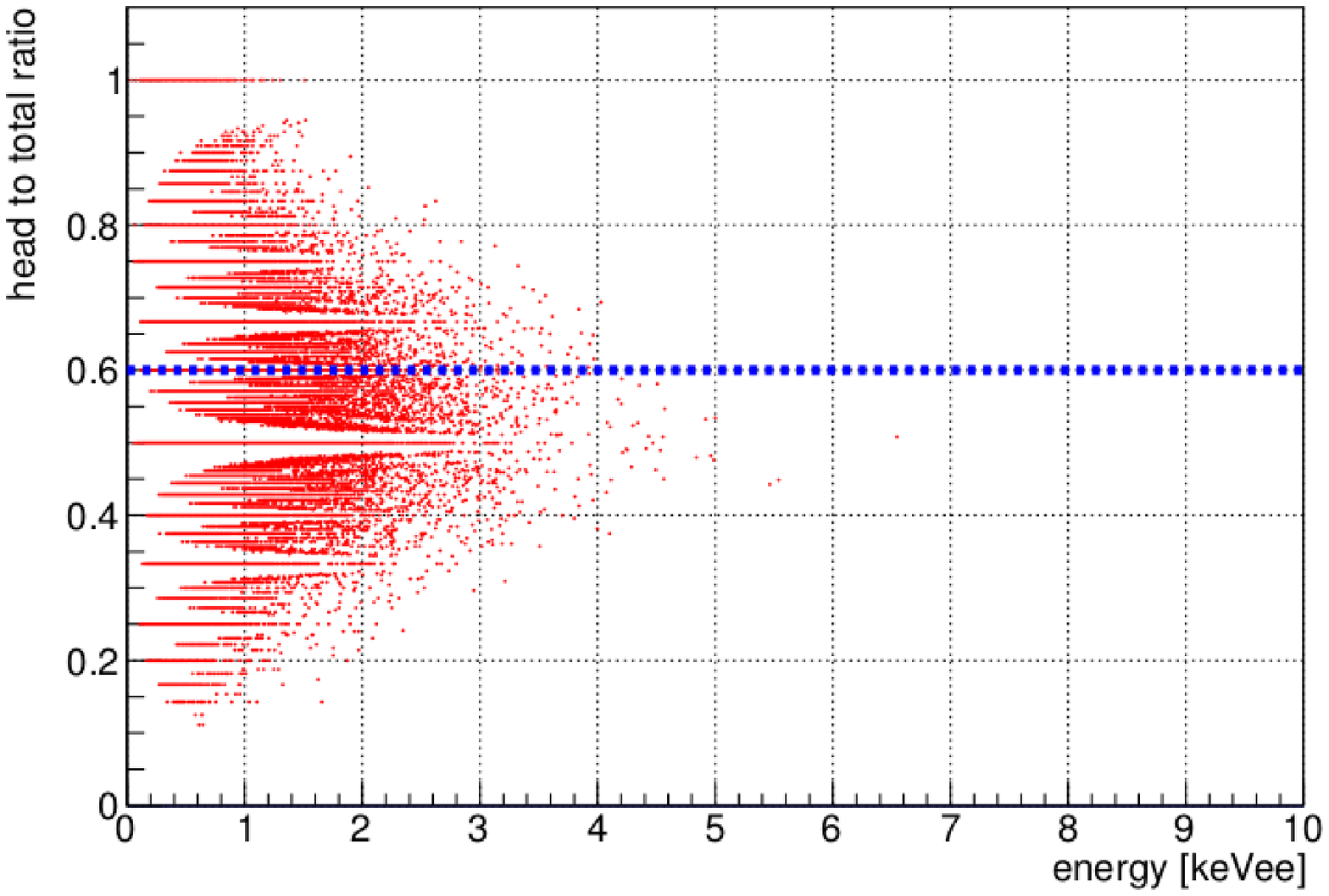}
\caption{The Cherenkov event (head-to-total ratio $> 0.6$) cut 
         for the XMASS data (left) and for simulated WIMP events (right) as a function of the observed energy.}
\label{fig:cherenkov}
\end{center}
\end{figure}

\begin{figure}[p]
\begin{center}
\includegraphics[width=8cm]{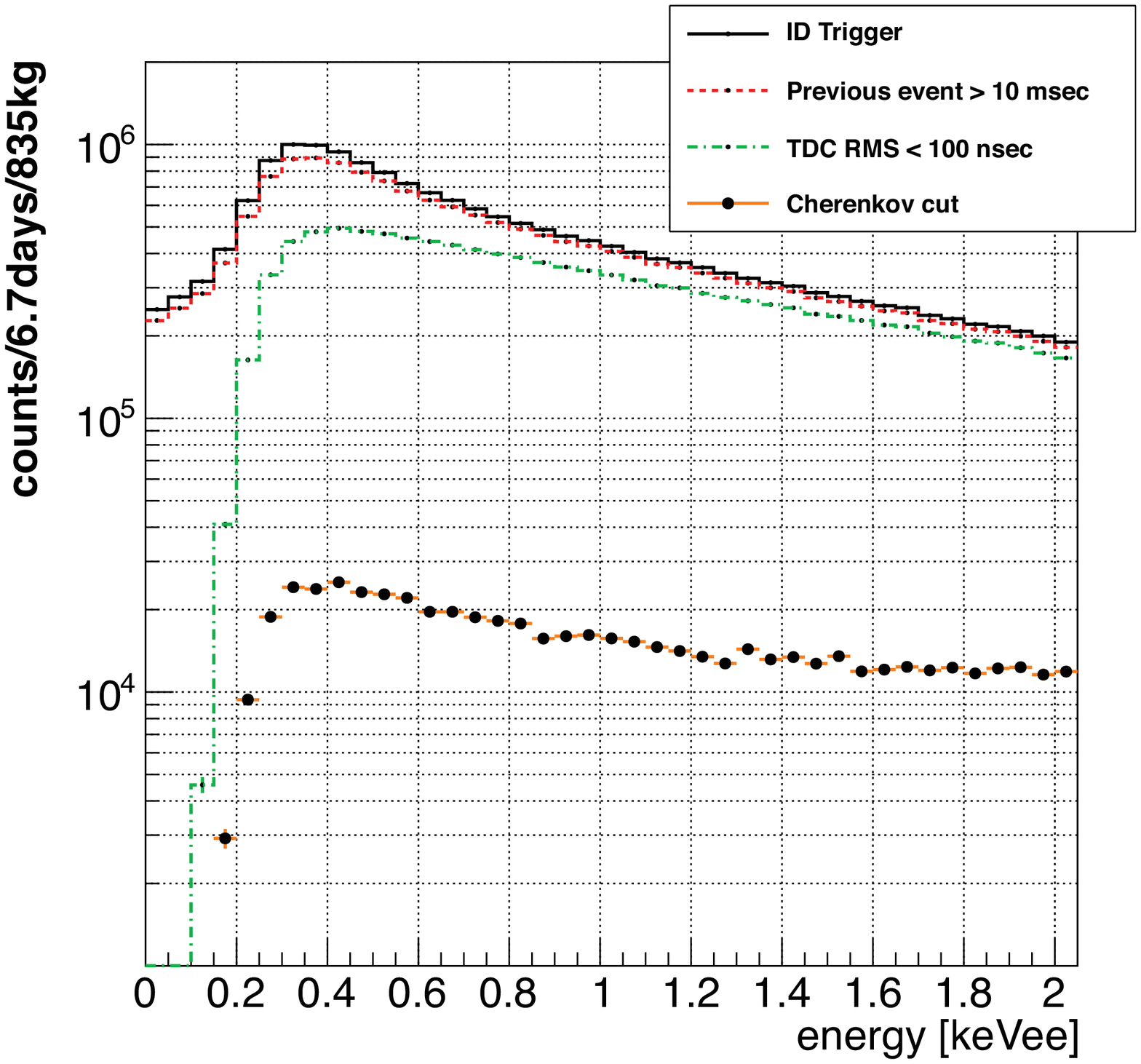}
\caption{The obtained energy spectrum at each step of the data reduction. 
         Raw data is shown as the solid line, the result of the 10\,ms to the previous event timing cut
         appears as the dashed line, and the dash-dotted line shows the data after application of the timing distribution
         width cut. The filled points show the result of the Cherenkov event cut. Details are presented in the text.}
\label{fig:reduction}
\end{center}
\end{figure}

The energy spectrum for all events in the entire volume of the detector over a 5591.4\,kg$\cdot$day exposure 
is shown in Fig.~\ref{fig:reduction}.
To remove backgrounds prior to analysis three types of timing cuts are applied to the raw data.
In order to remove events caused by the tail of the scintillation light distribution observed in 
energetic gamma, beta, alpha, and muon interactions within XMASS, events that occurred within 10\,ms 
of the previous event are rejected. 
Events whose timing distribution has 
an RMS greater than 100\,ns are also removed.  
The last and most significant cut is designed to remove events which produce light predominantly 
through Cherenkov emission, in particular the beta particle emitted in the decay of $^{40}$K 
contamination in the PMT photocathode.
To select these events a time-of-flight correction is made to the timing distribution of each event 
assuming the event vertex is on the surface of the PMT closest to the charge-weighted center of gravity of the event.
After this correction the timing distribution of Cherenkov-like events is found to be 
narrower than that for scintillation-like events. 
Events with more than 60\,\% of their PMT hits occurring within the first 20\,ns of the event window 
are removed as Cherenkov-like. The ratio of the number of 
PMT hits within the first 20\,ns relative to the total number of hits in the event window 
for all events (head-to-total ratio) is shown in Fig.~\ref{fig:cherenkov}.
Each step of the data reduction is shown in Fig.~\ref{fig:reduction}.
  
\begin{figure}[p]
\begin{center}
\includegraphics[width=8cm]{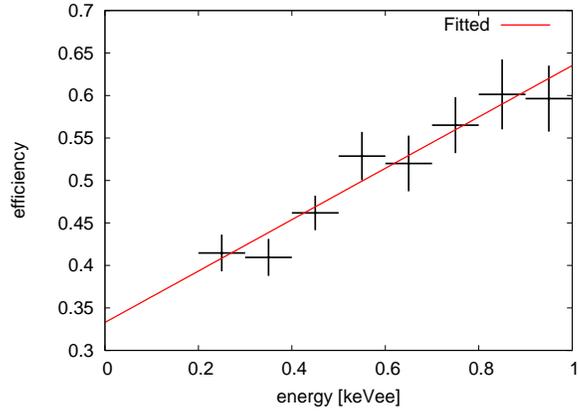}
\caption{WIMP signal acceptance efficiency after data reduction for the analysis.}
\label{fig:suveff}
\end{center}
\end{figure}

The expected WIMP acceptance efficiency of these cuts was estimated with the detector simulation.
In the simulation WIMP recoil energy spectra were generated for each WIMP mass and MC events were distributed uniformly 
throughout the detector volume using a liquid scintillation decay constant of 25\,ns~\cite{xmass_psd}.
Fig.~\ref{fig:suveff} shows the resulting signal acceptance efficiency at energies below 1\,keVee. 
The size of the error bars comes primarily from the systematic uncertainty in the   
xenon scintillation decay constant, $25\pm1$\,ns, which is estimated based on the difference
between the XMASS model~\cite{xmass_psd} and the NEST model~\cite{nest} based on~\cite{dawson}.
A systematic error on the selection efficiency is determined based on the error resulting from 
a linear fit to the points in the figure.  
At the 0.3\,keVee analysis threshold this error is 6.1\%.

\section{Results and Discussion}
\label{results}
\begin{figure}[p]
\begin{center}
\includegraphics[width=8.5cm]{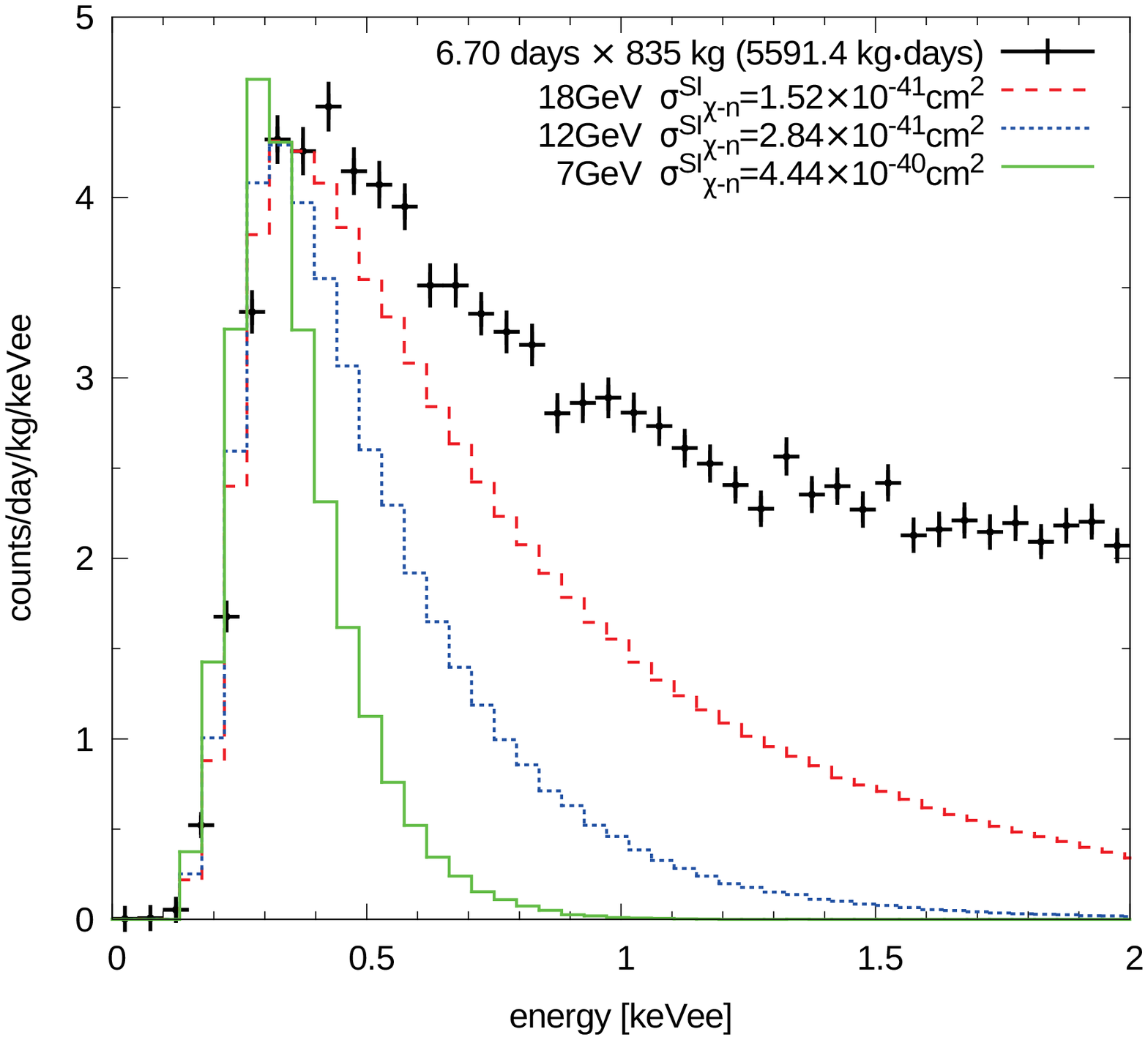}
\caption{Simulated WIMP energy spectra in the XMASS detector assuming the maximum cross section 
         that provides a signal rate no larger than the observation in any bin above 0.3\,keVee.}
\label{fig:spe}
\end{center}
\end{figure}

\begin{figure}[p]
\begin{center}
\includegraphics[width=8.5cm]{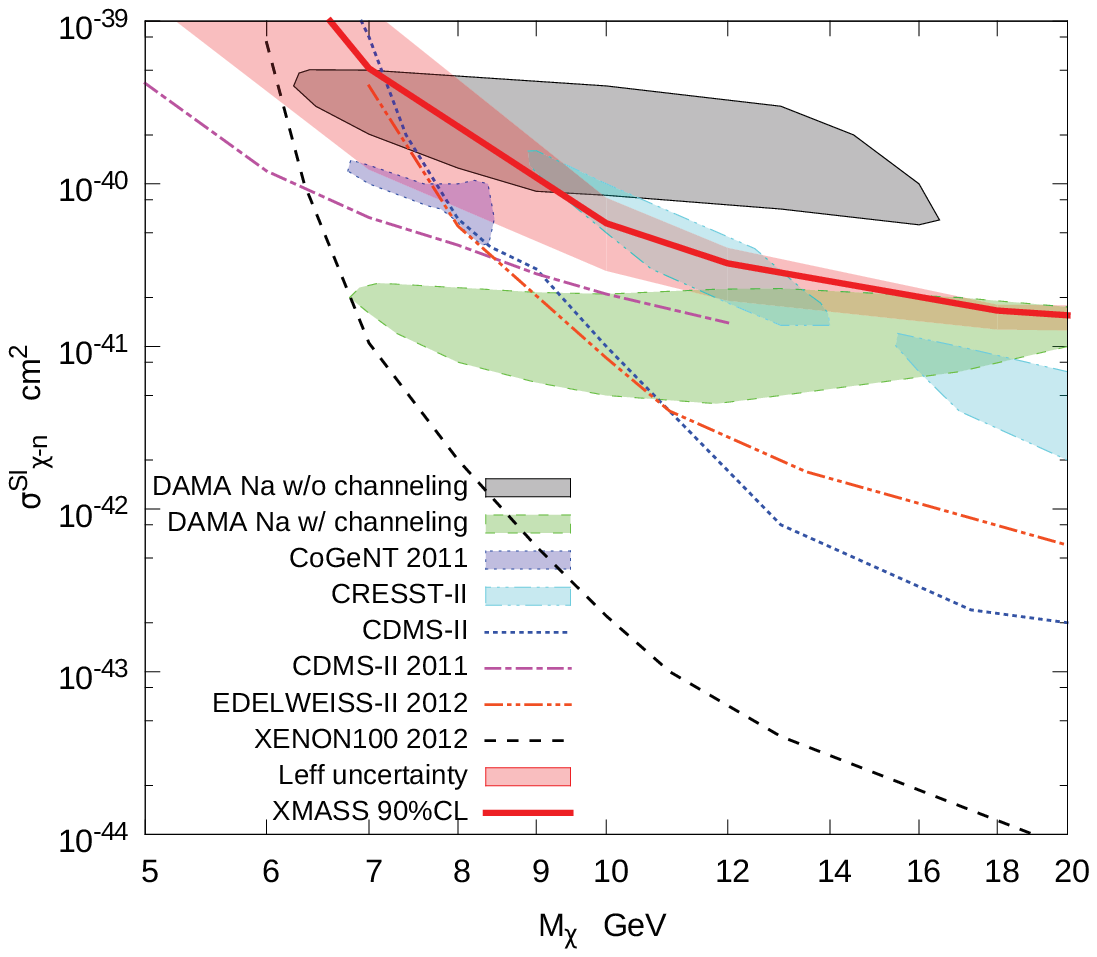}
\caption{Spin-independent elastic WIMP-nucleon cross section as a function of WIMP mass. 
         All systematic uncertainties except that from $\mathcal{L}_{\rm eff}$ are taken into account
         in the XMASS 90\,\% C.L. limit line. The effect of the $\mathcal{L}_{\rm eff}$ uncertainty 
         on the limit is shown in the band.
         Limits from other experiments and favored regions are also shown~\cite{dama, cogent, cresst, xenon, cdms, edelweiss}.} 
\label{fig:limit}
\end{center}
\end{figure}

Figure~\ref{fig:spe} shows simulated WIMPs energy spectra overlaid on the observed spectrum 
after the data reduction was applied. 
WIMPs are assumed to be distributed in an isothermal halo with $v_o=220$\,km/s, 
a galactic escape velocity of $v_{\rm esc}=650$\,km/s, and an average density of 0.3\,GeV/cm$^3$. 
In order to set a conservative upper bound on the spin-independent WIMP-nucleon cross section, 
the cross section is adjusted until the expected event rate in XMASS does not exceed the observed one 
in any energy bin above 0.3\,keVee.  
Implementing the systematic errors discussed in the text above, 
the resulting 90\,\% confidence level (C.L.) limit derived from this procedure is shown in Fig.~\ref{fig:limit}. 
The impact of the uncertainty from $\mathcal{L}_{\rm eff}$ is large in this analysis, so 
its effect on the limit is shown separately in the figure. 

After careful study of the events surviving the analysis cuts, their origins are not completely understood. 
Contamination of $^{14}$C in the GORE-TEX$^{\mbox{\scriptsize{\textcircled{\tiny R}}}}$ sheets between the 
PMTs and the support structure may explain a fraction of the events.
Light leaks through this material are also suspect. 
Nonetheless, the possible existence of a WIMP signal hidden under these and other backgrounds cannot be excluded. 
Although no discrimination has been made between nuclear-recoil and electronic events, 
and many events remain in the analysis sample, 
the present result excludes part of the parameter space favored by other 
measurements~\cite{dama,cogent,cresst} when those data are interpreted 
as a signal for light mass WIMPs. 
Finally, we are working on modifications to the inner surface of XMASS, 
especially around the PMTs, to improve the detector performance.


%

\section*{Acknowledgments}
We gratefully acknowledge the cooperation of Kamioka Mining and Smelting Company. 
This work was supported by the Japanese Ministry of Education, Culture, Sports, Science and Technology, Grant-in-Aid for Scientific Research, 
and partially by the National Research Foundation of Korea Grant funded by the Korean Government (NRF-2011-220-C00006).





\bibliographystyle{elsarticle-num}




\end{document}